# Nanoporous gold thin films as substrates to analyze liquids by cryo-atom probe tomography

Eric V. Woods[1,*], Aparna Saksena[1], Ayman A. El-Zoka[1,2], L.T. Stephenson[1,4], Tim M. Schwarz[1], Mahander P. Singh[1], Leonardo S. Aota[1], Se-Ho Kim[1,3], Jochen Schneider[5], Baptiste Gault[1,2,*]

1.  Max-Planck-Institut für Eisenforschung, Düsseldorf, Germany

2. Department of Materials, Royal School of Mines, Imperial College London, London, UK.

3. *now at* Australian Centre for Microscopy and Microanalysis, The University of Sydney, Sydney, NSW, Australia

4. *now at* Department of Materials Science and Engineering, Korea University, Seoul 02841, Republic of Korea

5. Materials Chemistry, RWTH Aachen University, Kopernikusstrasse. 10, 52074 Aachen, Germany

*Corresponding Authors: e.woods@mpie.de, b.gault@mpie.de

## Abstract

Cryogenic atom probe tomography (cryo-APT) is being developed to enable nanoscale compositional analyses of frozen liquids. Yet, the availability of readily available substrates that allow for the fixation of liquids while providing sufficient strength to their interface, is still an issue.  Here we propose the use of 1–2µm thick binary alloy film of gold-silver (AuAg) sputtered onto flat silicon, with sufficient adhesion without an additional layer. Through chemical dealloying, we successfully fabricate a nanoporous substrate, with open-pore structure, which is mounted on a microarray of Si posts by lift out in the focused-ion beam, allowing for cryogenic fixation of liquids. We present cryo-APT results obtained after cryogenic sharpening, vacuum cryo-transfer and analysis of pure water on top and inside the nanoporous film. We demonstrate that this new substrate has the requisite characteristics for facilitating cryo-APT of frozen liquids, with a relatively lower volume of precious metals. This complete workflow represents an improved approach for frozen liquid analysis, from preparation of the films to the successful fixation of the liquid in the porous network, to cryo-atom probe tomography.

# 1 Introduction

Atom probe tomography (APT) has emerged as an important method for compositional mapping on the nanoscale of solid materials with a sensitivity in the range of parts per million (Gault, et al., 2021). Especially in the biological area, one would benefit from the high local and especially chemical resolution from APT, since the structure of the proteins/organic molecules correlate directly with the functions and biochemical processes (Bellissent-Funel, et al., 2016). However, achieving this resolution requires the samples to be frozen and analyzed in their native state, and with APT requiring needle-shaped specimens with a radius of less than 100 nm, preparation has proved challenging(Schreiber et al., 2018; Stender et al., 2021). In recent years, progress has been made with the introduction of low-temperature and vacuum transfer shuttles as interest in the analysis of liquids in the atom probe has steadily increased (Gerstl & Wepf, 2015; Macauley, et al., 2021; Stender, et al., 2022; Stephenson, et al., 2018).

Considerable effort has been devoted to the development of different specimen preparation protocols to fix the liquid on or in a substrate and subsequently prepare suitable needle-shaped specimen by cryogenic focused ion beam (FIB) cutting the tip directly into a flat substrate with a water layer (El-Zoka, et al., 2020), embedding the liquid between graphene layers (Qiu, et al., 2020a; Qiu, et al., 2020b; Zhang, et al., 2022), infiltrating the liquid into a porous structure (Kim, et al., 2021), deposit a water droplet on a wire(Schwarz, et al., 2020), or cryo-lift-out (Schreiber, et al., 2018; Woods, 2023). Nevertheless, there is currently no standard protocol for liquid sample preparation, and different methods have their advantages and disadvantages. Some of the main remaining challenges are how to fix the liquid while providing sufficient strength to the interface between the substrate and the liquid, the control of the freezing conditions, minimizing the amount of liquid carried into the ultra-high vacuum chamber of the atom probe along with the preparation time.

In this paper, we introduce here an approach based on a thin film of AuAg deposited onto a Si wafer. Previous work has demonstrated the suitability of nanoporous gold (AuAg) as a substrate for APT (El-Zoka, et al., 2020). AuAg was obtained through de-alloying of a thin plate of a bulk alloy containing 77 % Ag and 23 % Au, which is a rather costly solution, and the use of a "moat" approach to prepare specimens requires a considerable amount of time for each individual needle.

We have successfully demonstrated that an AuAg thin film can be deposited via co-sputtering and that it adheres completely to the silicon substrate. Gold films often do not adhere well to silicon (Benjamin, 1997), so typically, using physical vapor deposition (PVD), an adhesion layer is needed (Volkov, et al., 2020), which, however, also influences the properties of the films (Todeschini, et al., 2017). Chromium as an adhesion layer would be attacked by the concentrated nitric acid used for AuAg chemical dealloying (El-Zoka, 2018; Seker, et al., 2009) leading to film delamination (Williams, et al., 2003; Williams & Muller, 1996). Titanium was hence preferred, and it is commonly used as an adhesion layer for gold PVD (Todeschini, et al., 2017) and is resistant to nitric acid corrosion (Corporation, 1997; Williams, et al., 2003; Williams & Muller, 1996). Since relatively thick sputtered thin films (e.g. microns) can have high stress gradients, it is critical to optimize deposition parameters to avoid delamination (Abadias, et al., 2018).

We use here room temperature lift-out in the focused-ion beam to prepare multiple lamellar pieces attached to the posts of a commercial support, that are subsequently made porous by chemical de-alloying in nitric acid to achieve nanoporous structure. Water was then injected into the porous structure in multiple specimens simultaneously using a straightforward immersion method. By freezing the sample immediately after, the liquid is fixated, and following annular milling, APT specimens were prepared and successfully investigated by APT. This cost-effective substrate and time-efficient method allowed a higher throughput and represents another possibility to improve APT analysis of liquid samples, especially of dissolved organic matter, in the future.

## 2 Materials & methods

### 2.1 Materials

Ag-Au thin films were grown onto 50 mm single crystal (001) silicon wafers, obtained from Siegert (Siegert Wafer GmbH, Aachen, Germany). For the support of lift-outs for APT analysis, a commercial 36 or 22 micropost coupon of conductively doped silicon microtips (FT 36/22, Cameca Instruments, Madison, WI, USA, hereafter "coupons") were utilized to mount specimens for APT analysis. The coupon was placed on Cameca copper specimen holders (hereafter "clips"), which were mounted on a Cameca cryogenic specimen puck, which have a special PEEK thermally isolating block to allow their transfer and temperature stability. An aluminium 12mm scanning electron microscope (SEM) stub and double-sided electrically conductive copper tape were used (Plano GmbH, Wetzlar, Germany), to fix the de-alloyed substrate to the stub for the lift-out. A piece of chromium (99.9% purity) was obtained from the MPIE metallographic workshop for redeposition welding.

### 2.2 Sputtered Thin Film Development

Ag-Au thin films were deposited in a combinatorial magnetron sputtering set-up, a schematic representation of which is shown in (Saksena, et al., 2018). All growth experiments were performed in an argon atmosphere at a pressure of 0.4 Pa. The base pressure before the depositions was ≤ 5 × $10^{-5}$ Pa. Individual elemental targets of Ag (99.9% purity) and Au (99.99% purity) were used. A film composition of $Au_{35}Ag_{65}$ was selected and deposition was realized ed by applying power densities of 6.1 W/cm$^2$ and 2 W/cm$^2$ to the Ag and Au targets, respectively. Simultaneously, the Si wafer is rotated to obtain a homogeneous composition at a growth temperature of 400 °C. These films are synthesized on top of (001) oriented Si wafer coated with a Ti adhesion layer, sputtered by applying a power density of 2.5 W/cm$^2$ where the wafer to target distance was maintained at 10 cm throughout the synthesis.

## 2.3 Initial sample preparation and chemical dealloying

Small pieces of AuAg-coated Si wafers were cleaved off the primary Si wafer using a diamond scribe. Each small piece was dropped into a small 10 mL beaker containing nitric acid for 30 seconds. Nitric Acid (69 %, Emsure ACS Reagent Grade fromMerck, Billerica, MA, USA) was used, removed using acid-resistant tweezers, and immediately rinsed in a beaker containing Type 1 ultra-pure water (UPW) twice and allowed to soak for a few minutes to exchange out any residual nitric acid.  Afterwards, the samples were blown dry using dry N2 gas.

## 2.4 Preliminary FIB lift out

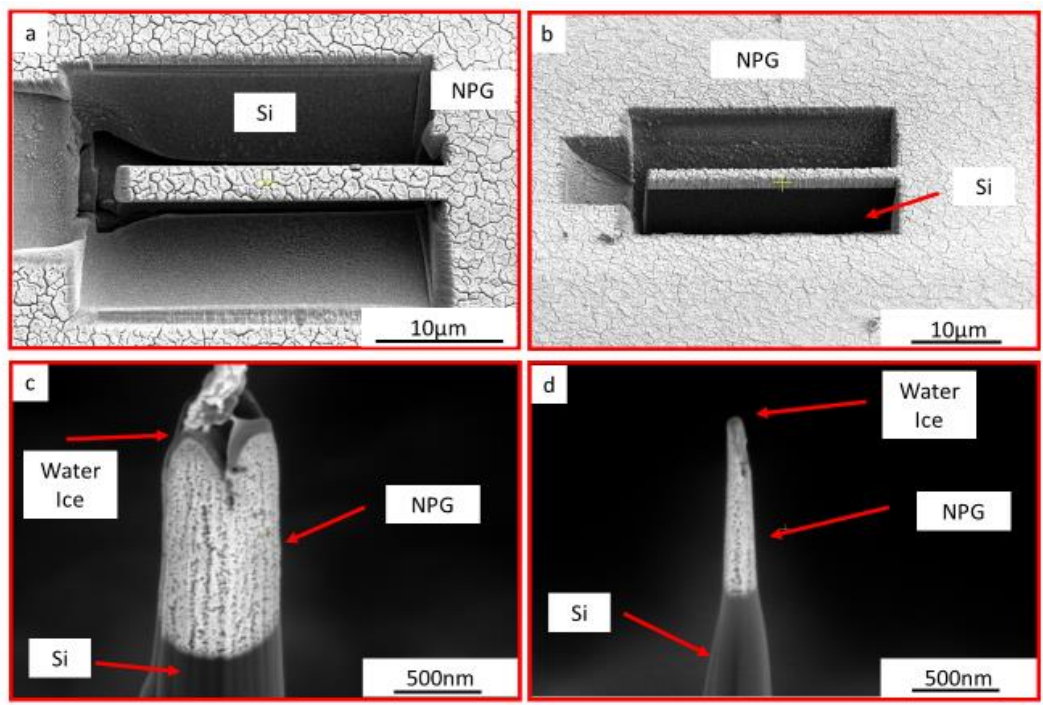

**Figure 1.** SEM images of AuAg water tip fabrication (a) Top view, APT lamella before liftout showing empty pores and cracks (b) Side view of APT lamella showing thickness (c) successful APT tip, fabricated from dipping into water, during sharpening illustrating thin water layer in much darker color on top, with a darker layer underneath, indicating filled pores (d) final sharpened APT tip

A Thermo Fisher Helios 5 CX Dual-Beam Ga FIB/SEM was used for RT (room temperature) sample preparation. All SEM images will be outlined in red, all Ion beam images in blue. The de-alloyed substrate was placed onto a 12 mm diameter Al SEM stub with double-sided Cu tape and loaded together with a coupon in a carousel holder. A lamella was prepared, approximately 23μm long, 2.5μm wide, and 5μm tall, as shown in **Figure *1*(a)**, following the protocol described in (Thompson, et al., 2007) and the chromium lamella preparation section of (Woods, 2023). Here, the bulk cuts were made with 30 kV, 0.79 nA ion beam current, and etched overnight to create minimal redeposition into the nanoporous AuAg. This was motivated by the impossibility to deposit a protective Pt layer on the surface, which would have filled the pores. The AuAg film thickness was measured to be approximately 2.5 μm, as illustrated in **Figure *1*(b).**

As a note, this approach was adopted because initial tests using cryo-liftout from a large piece of dealloyed AUAG that had bulk water on it, as in (Woods, 2023) were not as successful as desired, see **SI 2**.

Initially, to attach the extracted lamella onto a Si-post, a very small rectangle was used to deposit Pt at 30 kV, 40 pA ion beam current to minimize Pt deposition into the porous material. The gas-injection system (GIS) was then withdrawn to minimize the residual precursor gas in the chamber before the mounted piece was cut off from the lamella. A second set of data was obtained from specimens prepared using redeposition welding as outlined in (Woods, 2023).

During sharpening, the partially prepared porous AuAg film specimen showed a thin layer of water with a darker contrast on top as in **Figure *1*(e),** e.g. area of the AuAg film not excessively cracked, and visibly carrying liquid according to visual inspection in the SEM. The final sharpened tip is shown in **Figure *1*(f).** These will be further explained in the next section.

## 2.5 Preparation of Sample with Liquid and FIB sharpening

A Thermo Fisher Helios 5 CX Dual-Beam Ga FIB/SEM equipped with an Aquilos 2 freely rotating cryogenic stage, cryo EZ-Lift micromanipulator and a Ferroloader (Ferrovac GmbH, Zurich, Switzerland) vacuum transfer station, was used. The cryogenic stage was set to a target temperature of -190°C (with nitrogen gas flow of 190mg/sec), resulting in an actual average measured cryogenic stage temperature of -185°C and micromanipulator temperature of -175°C. The prepared coupon and clip were removed from the UHV chamber mounted on a cryo-puck (Cameca) and immediately transferred into the N2 glovebox (Sylatech) through a vacuum-cycled antechamber.

The puck was placed onto a brass holder such that the tips of the coupon were immersed in a glass beaker containing the liquid of interest (see **SI 1**). For the data in **Error! Reference source not found.**, **Error! Reference source not found.**, and **Error! Reference source not found.**, the coupon was left immersed in water for three minutes, and the data in **Error! Reference source not found.** the coupon was left immersed for thirty minutes. In both cases afterwards the puck was immediately plunged into liquid N2 to freeze and fix the water into the pores. The puck was then transferred into the pre-cooled Ferrovac D-100 UHV cryogenic Transfer Module (UHVCTM) suitcase.

The sample was transferred under vacuum at cryogenic temperature in the suitcase to the cryo-FIB, where it was loaded onto the pre-cooled cryogenic stage. During sharpening, the partially prepared porous AuAg film specimen showed a thin layer of water with a darker contrast on top, as in **Figure 1(e).** An area of the AuAg film not excessively cracked, and visibly carrying liquid according to visual inspection in the SEM, was targeted for making the APT specimen. In the upper area of the AuAg tip, an area of darker contrast directly below the water layer is visible, which suggests that the pores there are filled with water. The specimen was sharpened with decreasing currents and decreasing inner diameter until a final diameter of approx. 100 nm was achieved, as shown in **Figure 1(f).** Since Ga-beam damage under cryogenic conditions is deemed negligible, no low-voltage cleaning step was performed (Bassim, et al., 2012; Lucas & Grigorieff, 2023; Parkhurst, et al., 2023; Parmenter, et al., 2014). Afterwards, the puck was transferred back into the UHVTM suitcase using the Ferroloader.

## 2.6  Atom Probe Analysis

A Cameca LEAP 5000XR reflectron (Cameca Instruments Inc.) equipped for cryogenic sample transfer (Stephenson, et al., 2018) was used for data acquisition. Data was gathered using the following measurement parameters for laser-pulsing mode: 50K, 75-100 pJ laser pulse energy, 125 kHz pulse repetition rate, 0.005 ions per pulse average detection rate (**Error! Reference source not found.**, **Error! Reference source not found.**). For voltage-pulsing mode, the followed parameters were used: 50 K, 15% pulse fraction, 125 kHz pulse rate and 0.005 ions per pulse detection rate (**Error! Reference source not found.**).  For the redeposition-welded material in laser mode (**Error! Reference source not found.**), run conditions were: 60K, 125kHz, 100 pJ, 0.005 ions per pulse detection rate.

# 3   Results

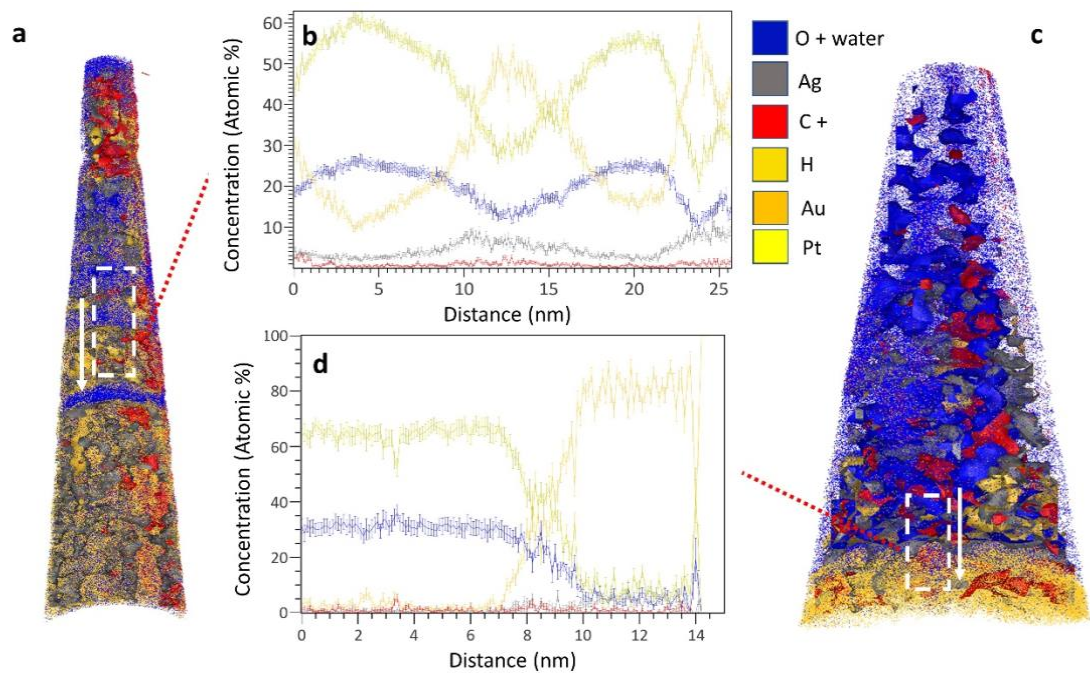

**Figure 2**. Water-filled nanoporous gold with organic contamination from Pt precursor (a) Laser mode dataset with (b) inset region of interest showing water filled gold pores (c) voltage mode dataset with (d) inset region of interest showing water

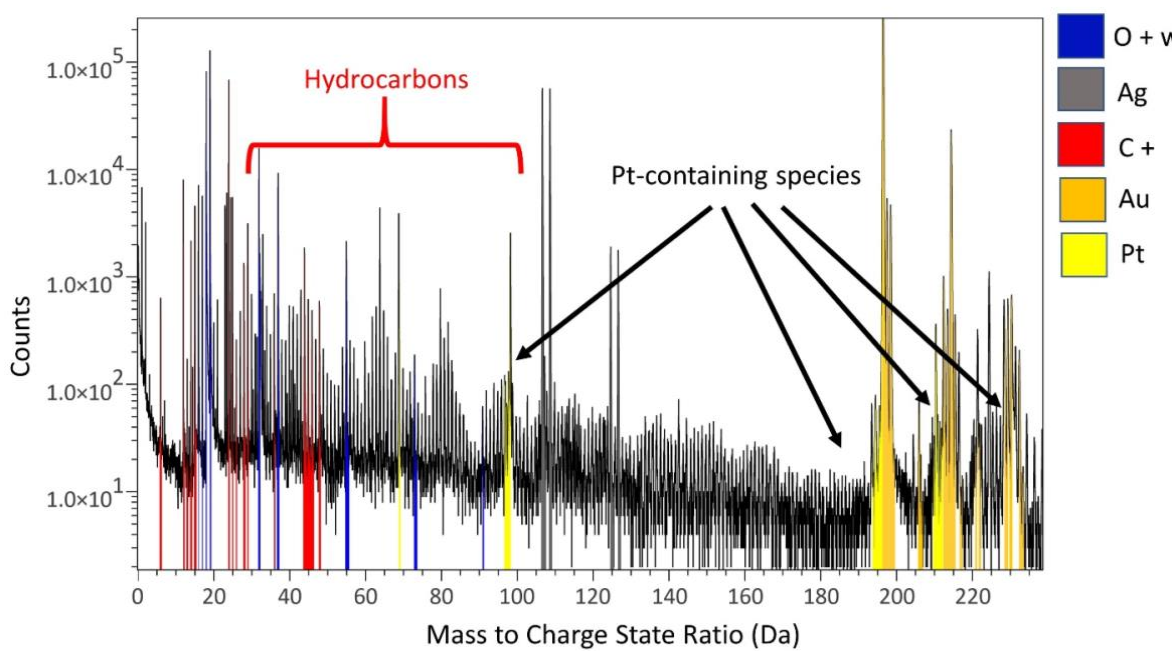

**Figure 3**. Laser mode mass spectra showing carbonaceous and platinum-containing species contamination derived from platinum precursor

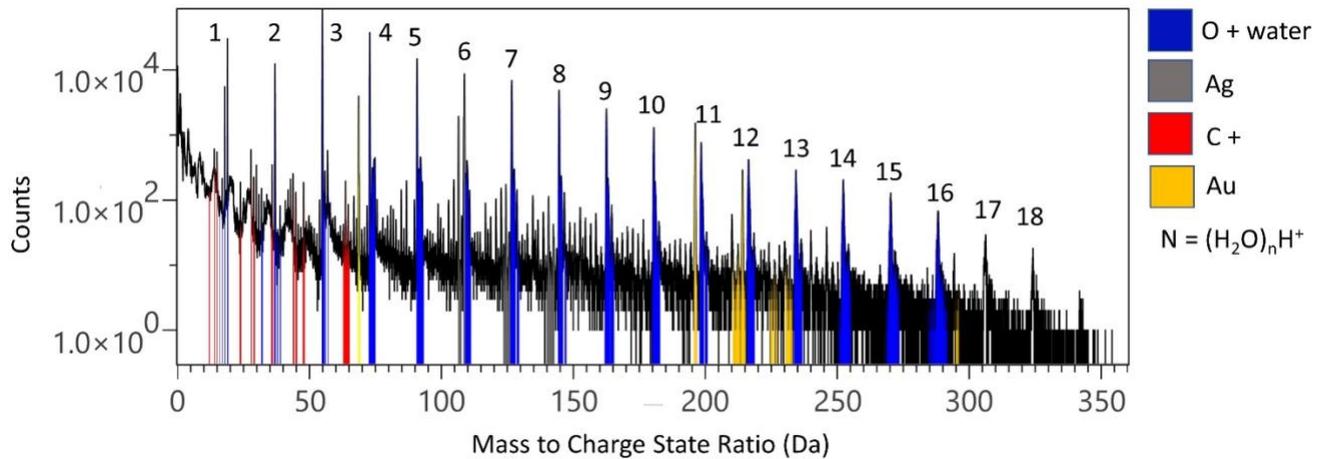

**Figure 4.** Low-field mass spectra demonstrating water clusters to n=18

**Error! Reference source not found.** summarizes data obtained both in laser and voltage pulsing mode, and containing water and the interface with the substrate. As expected, silver was largely removed through etching and the water was present in gold pores. As **Error! Reference source not found.** clearly illustrates, there was significant contamination by carbon-containing species (denoted C+) indicated in red, which make it extremely difficult to identify any species of interest not at substantial abundance. Voltage mode reconstructions were used for laser mode data. A typical laser mode reconstruction is shown in **Error! Reference source not found.(a)** with an inset region of interest (ROI) in **Error! Reference source not found.(b)** showing pore-filling. A typical voltage mode reconstruction is shown in **Error! Reference source not found.(c)** with an interfacial ROI in **Error! Reference source not found.(d)**, showing fewer pores in this structure, which uses a fixed shank reconstruction, which gives an indicative shape and structure. A typical laser mode mass spectrum is shown in **Figure 3** illustrates the innumerable peaks showing organic species. Given that the AuAg was soaked in ultrapure Type I water, the organics must have come from residual organometallic precursor that infiltrated and remained in the pores, and later dissolved into the water and decomposed. Interestingly, as demonstrated in previous work (Woods, et al., 2023) showing that water can show clusters in voltage mode APT out to a cluster number of n=18, under apparently low field conditions the same phenomenon can be observed in laser mode, shown in **Figure 4**. This demonstrates that preservation of high-molecular weight analytes of interest is possibly in either laser or voltage mode pulsed APT. These initial results validated the AuAg thin film as substrate for analyzing liquids containing analytes of interest, yet GIS-free attachment was demonstrably necessary to avoid carbon contamination.

Pure water was then analyzed by using redeposition welding, i.e. avoid the GIS altogether, as shown in the 3D reconstruction in **Error! Reference source not found.(a)** and the overall mass spectrum in **Error! Reference source not found.(b).** In a sample nanoporous interface region showing a 1D cross-sectional elemental profile in **Error! Reference source not found.(c),** the crossover between water and the surface and partially filled pore is illustrated.

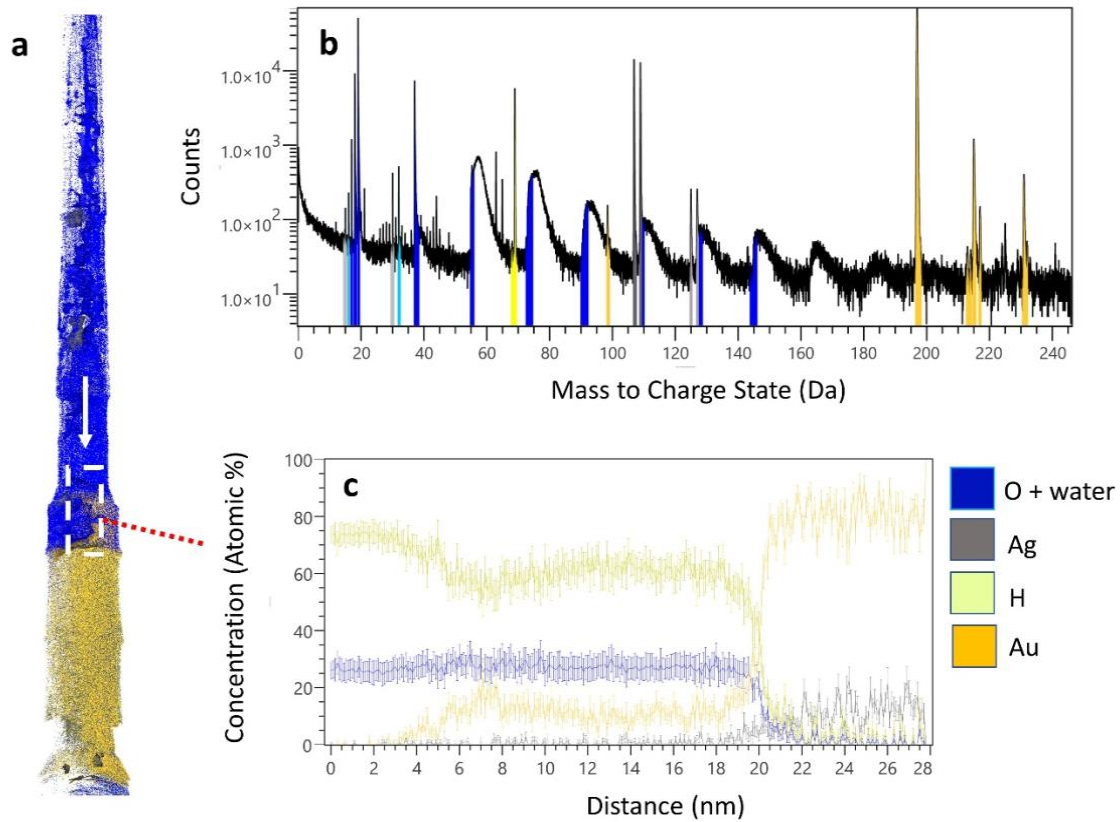

**Figure 5.** Redeposition-welded AuAg with 30 minutes immersion (a) overall reconstruction with ROI shown (b) Overall mass spectrum showing no carbon species (c) Zoomed ROI showing Z-axis 1D composition profile

# 4 Discussion

## 4.1 Why Thin Films are an Important Advancement for cryo-APT

The use of a thin film of AuAg alleviates the need for bulk AuAg (El-Zoka, 2018; El-Zoka, et al., 2017). Previous work on AuAg (El-Zoka, et al., 2020) reported on specimens prepared by using the "crater" (Halpin, et al., 2019; Miller, et al., 2005) or "moat" (El-Zoka, et al., 2020) method. Of note, that method limits sample fabrication to the edge of a bulk specimen and requires extensive milling of large quantities of material and significant FIB and personnel time, thus substantively limiting throughput. We have demonstrated a simple sample preparation protocol for dealloying the AuAg thin film and that pieces can be lifted out and mounted to a Si-post at room temperature, increasing throughput, since half a dozen tips can be made and used later. We have demonstrated that a Pt GIS can be used to attach dealloyed AuAg pieces to silicon microtips, but that such attachment procedure infiltrates the pores and creates substantial organic contamination. Redeposition welding (Douglas, et al., 2022; Schreiber, et al., 2018; Woods, et al., 2023) allows for successful specimen preparation for analytic purposes with no resultant contamination. Notably, this method allows laboratories with older non-rotating cryogenic stages to utilize this technique.

While the approach we outlined herein could be used on a lifted-out piece from the bulk AuAg alloy, subsequently subjected to dealloying. However, the nanoporous structure could then extend all the way into the redeposition weld, which was made of the lamella material in this paper, along with Ga (Woods, et al., 2023). However, during dealloying in nitric acid, the Ag-containing weld itself was weakened, and lead to facile fracture during APT analysis. In the thin film geometry, the passive Ti-interfacial region ensures that the dealloying process is limited to the AuAg layer, and that the weld is of Si-on-Si, making it strong as demonstrated by the possibility of running in high-voltage pulsing mode as described in (Woods, et al., 2023).

## 4.2 Applications

This method allows for high-throughput production of liquid-containing atom probe tips. Provided that the chemical dealloying process is optimized to remove any silver, it also provides a framework material which is chemically inert and has a reasonably good electrical and thermal conductivity. As we have demonstrated in other work (Woods, et al., 2023), aqueous solutions containing organic analytes can be successfully analysed.

## 4.3 Consideration for future work

With respect to the sputtered thin films, we could explore alternative routes to increase adhesion such as ion bombardment (Guzman, et al., 2002; Martin, 1986) and other types of surface treatment (Li, et al., 2004),or micro-sputtering (e.g. additive-manufacturing) techniques (Kornbluth, et al., 2020), or deposition at higher temperature, which helps control parameters including homogeneity and stress (Chauvin, et al., 2020; Saksena, et al., 2018).

We have demonstrated that the water-filled AuAg structure can be successfully measured by cryo-APT, thanks to open pores network of the de-alloyed AuAg film making the surface hydrophilic (Abdelsalam, et al., 2005; El-Zoka, 2018; Yokomaku, et al., 2008). The successful fixation of little volumes of frozen fluids on top of Si-post in this paper, could very well pave the way for a very simple, straightforward method for future cryo-APT analyses of frozen liquids. A caveat to the method is that the porous material can be only partially filled with liquid in the top region, which can be optimized by a longer immersion time. To increase the infiltration of the liquid into the AUAG structure, a longer immersion time can be used and plasma cleaning the AuAg specimen immediately before use to improve hydrophilicity (e.g. contact angle) could also be tried to increase the thickness of the retained water layer. For increasing the throughput, the AuAg thin films could be deposited directly onto a commercial flat-top Si-post microtip array (Hans, et al., 2021), which, following dealloying, would allow to infiltrate numerous samples at the same time . Finally, biologically-relevant or chemically-relevant solutions may not infiltrate as easily as water, making their preparation and analysis more difficult, thereby requiring optimisation of pore size and immersion time.

# 5 Conclusion

To summarize, magnetron sputtering has been used to synthesize AuAg thin films. A protocol for using these as a carrier for analysis of liquids in the APT was then proposed. Following dealloying in nitric acid at room temperature, we performed lift-out in the FIB and pre-shaped the support into cylinders with a diameter below 1 micron. Immersion in liquid and plunge freezing into liquid nitrogen allowed for a simple cryogenic transfer step to the FIB for sharpening has been demonstrated.  The use of redeposition welding prevented ingress of the gaseous precursor from the gas-injection system into the nanoporous network. We have successfully demonstrated this complete simplified sample preparation wokflow for cryo-APT.

# 6 Conflict of Interest Statement
The authors declare no conflicts of interest.

# 7 Acknowledgements and Funding

EVW, AEZ, SHK, and BG are grateful for funding from the European Research Council (ERC) for the project SHINE (ERC-CoG) #771602. BG and TMS is grateful to the Deutsche Forschungsgemeinschaft (DFG) for funding through the Leibniz Award.  AS is grateful for funding through.  LS is grateful for funding through.  We thank Uwe Tezins, Christian Broß and Andreas Sturm for their support at the FIB and APT facilities at MPIE. We would like to thank Katja Angenendt, Monika Nellessen, and Christian Broß for their assistance with sample preparation.

# 9 Supplementary Information

**SI 1.** Experimental setup for immersing APT tips in liquid. A small glass bowl was placed on a laboratory stand and raised so that only the AuAg tips on the coupon were immersed into the liquid meniscus.

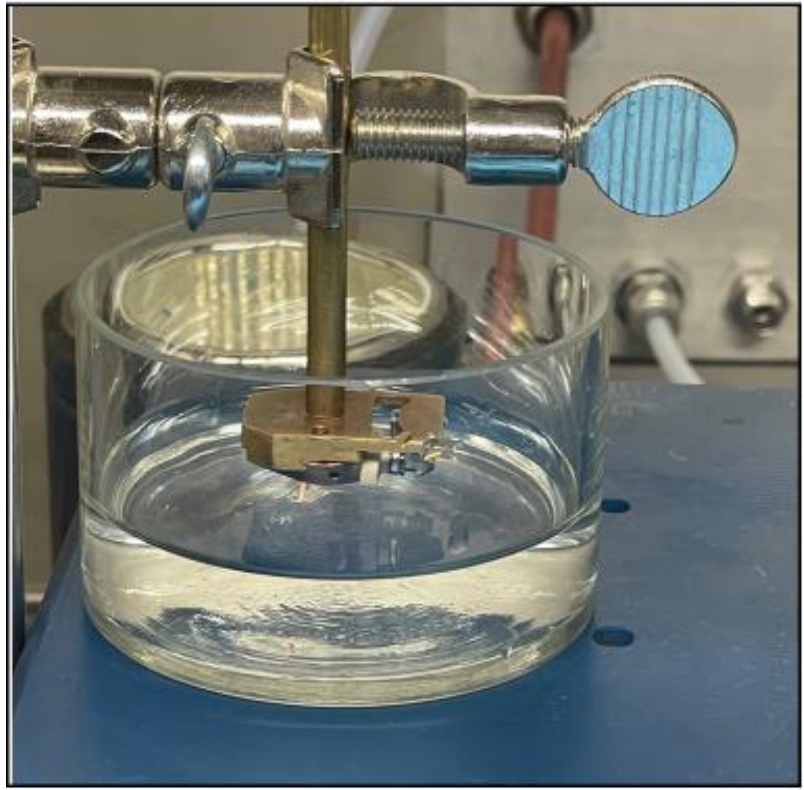

*SI 2. Cryogenic liftout of NPG thin films: (a) cross-section of lamella before liftout and (b) backscattered electron image of NPT ip with water ice)*

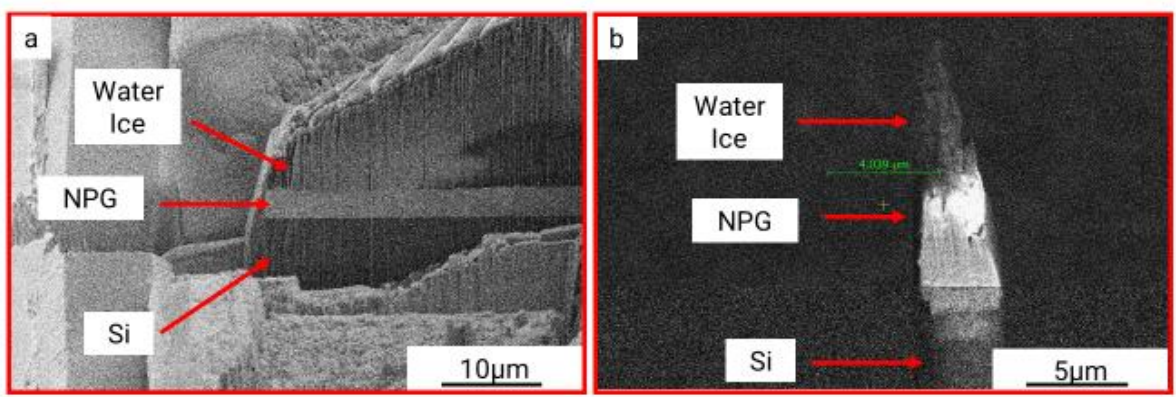

(a) initial bulk water on AuAg film (b) APT tip from bulk water showing lack of water integrity, with clear contrast of the water layer on top, and a lighter contrast in the pores indicating gold